\documentclass[aps,prl,twocolumn,floatfix,showpacs,citeautoscript,superscriptaddress,longbibliography,hyperlinks]{revtex4-2}
\usepackage{comment}
\usepackage[T1]{fontenc}
\usepackage{amsmath, amsfonts, amssymb}
\usepackage{tabularx, graphicx, xcolor}
\usepackage{siunitx}
\usepackage[colorlinks=true,citecolor=blue]{hyperref}
\usepackage{bbold, bm, bbm} 
\usepackage{tikz}
\usepackage{enumitem}
\usepackage{lineno}
\usepackage{svg}

\newcommand{\md}{\mathrm{d}}

\newcommand{\me}{\mathrm{e}}

\definecolor{magenta2}{RGB}{255,102,255}
\definecolor{crimson}{RGB}{220,20,60}

\usepackage[capitalise]{cleveref}

\renewcommand{\vec}[1] {\mathbf{#1}}
\newcommand{\up}{\uparrow}
\newcommand{\dw}{\downarrow}

\newcommand{\sid}{\sigma_0}
\newcommand{\sx}{\sigma_x}
\newcommand{\sy}{\sigma_y}
\newcommand{\sz}{\sigma_z}

\newcommand{\tid}{\tau_0}
\newcommand{\tx}{\tau_x}

\newcommand{\tz}{\tau_z}

\usepackage{mathtools}

\newcommand{\ket}[1] {\lvert #1{\rangle}}

\newcommand{\braket}[2] {{\langle #1{\lvert} #2{\rangle}}}

\usepackage{glossaries}
\glsdisablehyper
\newacronym{tsc}{TSC}{topological superconductivity}
\newacronym{2deg}{2DEG}{two dimensional electron gas}
\newacronym{mzm}{MZM}{Majorana zero modes}
\newacronym{car}{CAR}{crossed Andreev reflection}
\newacronym{ec}{EC}{electron cotunneling}
\newacronym{ll}{LL}{Landau Level}
\newacronym{caes}{CAES}{chiral Andreev edge state}
\newacronym{rsoc}{RSOC}{Rashba spin-orbit coupling}
\newacronym{pbc}{PBC}{periodic boundary conditions}
\newacronym{sc}{SC}{superconductor}
\newacronym{bdg}{BdG}{Bogoliugov de Gennes}
\newacronym{qh}{QH}{quantum Hall}
\newacronym{nn}{NN}{nearest-neighbours}
\newacronym{l}{$L$}{left}
\newacronym{r}{$R$}{right}

\usepackage{sidecap,tikz}
\definecolor{lime}{HTML}{A6CE39}
\DeclareRobustCommand{\orcidicon}{\hspace{-1mm}
	\begin{tikzpicture}
		\draw[lime, fill=lime] (0,0) 
		circle [radius=0.16] 
		node[white] {{\fontfamily{qag}\selectfont \tiny \,ID}};
		\draw[white, fill=white] (-0.0525,0.095) 
		circle [radius=0.007];
	\end{tikzpicture}
	\hspace{-3mm}
}
\foreach \x in {A, ..., Z}{\expandafter\xdef\csname orcid\x\endcsname{\noexpand\href{https://orcid.org/\csname orcidauthor\x\endcsname}
		{\noexpand\orcidicon}}
}

\newcommand{\uam}{Department of Theoretical Condensed Matter Physics\char`,~Universidad Aut\'onoma de Madrid, 28049 Madrid, Spain}
\newcommand{\ifimac}{Condensed Matter Physics Center (IFIMAC), Universidad Aut\'onoma de Madrid, 28049 Madrid, Spain}
\newcommand{\inc}{Instituto Nicol\'as Cabrera, Universidad Aut\'onoma de Madrid, 28049 Madrid, Spain}

\begin{document}

\title{Emergent topology by Landau level mixing in quantum Hall-superconductor nanostructures}

\author{Yuriko Baba\orcidA{}}
\email{yuriko.baba@uam.es}
\affiliation{\uam}
\affiliation{\ifimac}

\author{Alfredo Levy Yeyati\orcidB{}}
\email{a.l.yeyati@uam.es}
\affiliation{\uam}
\affiliation{\ifimac}
\affiliation{\inc}

\author{Pablo Burset\orcidC{}}
\email{pablo.burset@uam.es}
\affiliation{\uam}
\affiliation{\ifimac}
\affiliation{\inc}

\date{\today}

\begin{abstract}
We demonstrate the emergence of novel topological phases in quantum Hall-superconductor hybrid systems driven by Landau level mixing and spin-orbit interactions. 
Focusing on a narrow superconducting stripe atop a two-dimensional electron gas, we identify regimes where the hybridization of the chiral Andreev states at each side of the stripe leads to different phases beyond the long sought $p$-wave superconducting one. These topological phases exhibit distinctive transport signatures, including quantized nonlocal conductance arising from electron cotunneling at filling factor $\nu=1$, which can coexist with quantized crossed Andreev reflection at $\nu=2$. A combination of numerical simulations and effective modelling reveals the role of spin-orbit coupling and stripe geometry in controlling these transitions. Our findings suggest new strategies for realizing and detecting topology in proximized quantum Hall devices.
\end{abstract}

\maketitle

%\section{Intro}

\emph{Introduction.}---
\Gls{qh}-\gls{sc} hybrid structures provide an attractive platform in the search of engineered topological phases \cite{Clarke2014,Mong2014, Qi2010}. In particular, $p$-wave topological superconductivity is expected to arise when two counter-propagating spin-polarized edge modes are coupled by a conventional \gls{sc}~\cite{Fu2008,Qi2010,Finocchiaro2018, Galambos2022}. 

While these proposals are based on idealized conditions, recent experimental progress in realizing proximized \gls{qh} devices~\cite{Amet2016, Park2017, Lee2017,  Wei2019,Seredinski2019, Sahu2021,
Guel2022,Zhao2023,Vignaud2023,MehdiHatefipour2023,Hatefipour2024, Akiho2024} has boosted theoretical investigations beyond early models~\cite{Hoppe2000, Giazotto2005, Akhmerov2007, Khaymovich2010}. These studies explore the influence of the interface transparency~\cite{ Bondarev2025}, disorder~\cite{Manesco2022,Kurilovich2023,Kurilovich2023a, Cuozzo2024}, device geometry~\cite{David2023, Beconcini2018, Khrapai2023}, magnetic field penetration~\cite{Michelsen2023, Galambos2022}, the effect of vortices~\cite{Zocher2016,Tang2022,Kurilovich2023,Schiller2023}, or heat transport~\cite{Panu2024}. 
Additionally, other transport signatures like shot noise have been explored~\cite{Gamayun2017, Arrachea2024}. Despite this progress, a definitive experimental confirmation of the presence of topological superconductivity in \gls{qh} devices is lacking. 

\begin{figure}[b]
    \centering
    \includegraphics[width=\linewidth]{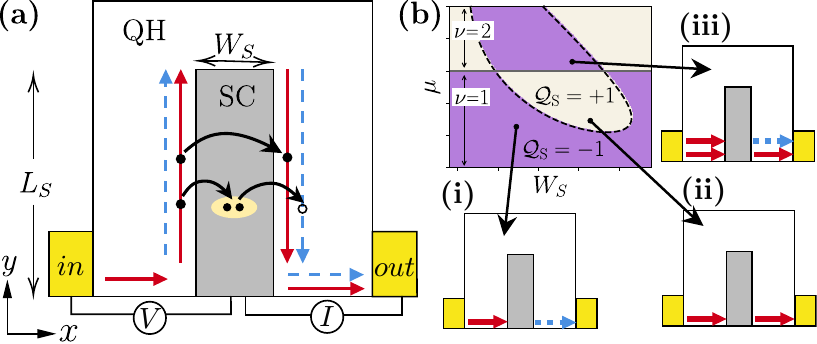}
    \caption{(a) Sketch of a \gls{2deg} in the \gls{qh} regime with a superconducting finger of width $W_S$ placed on top. 
    % The \gls{2deg} chemical potential $\mu$ can be varied by external gates. 
    Solid red and dashed blue arrows respectively represent electron and hole edge states coupled by the \gls{sc} (gray). 
    In a nonlocal transport measurement, voltage is applied bewteen lead $in$ and \gls{sc} and current is measured in lead $out$. 
    (b) Schematic phase diagram in the $\mu-W_S$ plane for an infinite \gls{sc} stripe. Panels (i-iii) indicate the dominant transport processes in each region of the phase diagram for a finite-length \gls{sc} finger. }
    \label{fig:1Sketch}
\end{figure}

In this work we consider a setup where a narrow superconducting finger is placed on top of a Hall bar geometry defined on a \gls{2deg}, see \cref{fig:1Sketch}(a). 
% \al{As discussed in Refs.~\cite{Clarke2014, Finocchiaro2018, Galambos2022}, the emergence of topological superconductivity in such devices for $\nu = 1$ would manifest in a negative quantized nonlocal conductance at subgap voltages, originating from resonant Andreev reflections at the Majorana zero modes localized at the ends of the superconducting finger. }
% \pb{As discussed in Refs.~\cite{Clarke2014, Finocchiaro2018, Galambos2022}, the emergence of topological superconductivity in such devices for $\nu = 1$ would manifest with the formation of Majorana zero modes at the top and bottom terminations of the finger [\cref{fig:1Sketch}(a)]. Consequently, resonant Andreev processes at these Majorana states induce a negative quantized nonlocal conductance at subgap voltages. }
As discussed in Refs.~\cite{Clarke2014, Finocchiaro2018, Galambos2022}, topological superconductivity emerges for $\nu = 1$ when the finger width $W_S$ is comparable to the superconducting coherence length forming Majorana zero modes at its top and bottom terminations. 
In a nonlocal transport configuration like the one in \cref{fig:1Sketch}(a), topological superconductivity would manifest in perfect \gls{car}, i.e., perfect electron-hole conversion as depicted in panel (i) of \cref{fig:1Sketch}(b). 
%This behaviour corresponds to perfect \gls{car}, i.e., perfect electron-hole conversion as depicted in panel (i) of \cref{fig:1Sketch}(b). 
%
%By contrast, for a set of parameters outside the topological phase no such quantization is expected and the nonlocal conductance $G_{\rm nl} = (e^2/h)(T_{\rm EC} - T_{\rm CAR})$ could take any value (positive or negative), as a result of the competition between \gls{car} and \gls{ec} processes. 
%\pb{define nonlocal, , inducing perfect \gls{car}, i.e., perfect electron-hole conversion as depicted in panel (i) of \cref{fig:1Sketch}(b). }
By contrast, outside the topological phase no such quantization is expected and the nonlocal conductance 
$G_{\rm nl} = (e^2/h)(T_{\rm EC} - T_{\rm CAR})$ could take any value (positive or negative), as a result of the competition between \gls{car} and \gls{ec} processes. 

We demonstrate that other previously unnoticed topological phases, with drastically different transport properties, can arise for \gls{qh}-\gls{sc} hybrid systems. These phases feature quantized transport channels with perfect \gls{ec} and emerge when spin-orbit coupling dominates over Zeeman splitting and for pairs of \glspl{ll} with an energy separation comparable to the superconducting gap. % $\Delta$. 

A schematic phase diagram for an infinitely long \gls{sc} stripe with periodic boundary conditions along the $y$ direction is presented in \cref{fig:1Sketch}(b). 
Pairs of electron and hole \glspl{ll} at each edge of the stripe form \glspl{caes} and the inter-edge hybridization between them defines a set of subbands as a function of the parallel momentum $k_y$. 
When the chemical potential $\mu$ approaches the boundary between adjacent filling factors, and depending on the stripe width $W_S$, a band inversion can occur marked by a sign change in the associated Pfaffian invariant ${\cal Q}_\mathrm{S}$. 
In those regions [brighter areas in \cref{fig:1Sketch}(b)] and for $\nu = 1$, the nonlocal conductance in the finger geometry would jump to a quantized {\it positive} value, leading to perfect \gls{ec} transmission [panel (ii)]. Remarkably, when increasing $\mu$ to reach $\nu=2$ within these regions, the nonlocal conductance completely vanishes due to the opposite quantized contributions by \gls{ec} and \gls{car} processes corresponding to the two edge modes [panel (iii)]. 

In the first part of this work we analyze the phase diagram in terms of relevant parameters for the stripe geometry. We combine numerical calculations with an effective analytical model which allows us to get a simple physical picture of the origin of the different topological phases. We then present results for the transport properties which confirm the expected quantization for each phase, as well as their robustness against several factors. 

\emph{Model and phase diagram analysis.}---
The system considered is a proximized \gls{2deg} in the \gls{qh} regime controlled by a perpendicular magnetic field $B_z$. 
The corresponding model Hamiltonian can be expressed as
\begin{align} \label{eq:model:ham}
    \mathcal{H} & = \left( \frac{\hbar^ 2 \vec{k}^ 2}{2 m}
    -\mu(\vec{r}) \right) \sid \tz 
    + V_z (\vec{r}) \sz \tid
     \nonumber \\ 
    &  + \alpha_R(\vec{r}) \left(k_x \sy \tid - k_y \sx \tid \right) - \Delta (\vec{r}) \tx \sz~,
\end{align}
where $\vec{k} = -i\vec{\nabla}_\vec{r}$, while $\sigma_j$ and $\tau_j$, with $j = x, y,z$, are Pauli matrices corresponding to the spin and electron-hole spaces, respectively. 
The orbital part of the magnetic field is included by the minimal substitution $\hbar \vec{k} \to \hbar \vec{k} - q \vec{A}(\vec{r})$, where $\vec{A}(\vec{r})$ is the vector potential. 
This model includes the \gls{rsoc} through the $\alpha_R(\vec{r})$ term, the proximity-induced $s$-wave pairing ($\Delta(\vec{r})$ term)
and the Zeeman splitting $V_z(\vec{r})  = g \mu_B B_z(\vec{r})/2$. 
We consider the \gls{rsoc} to be constant throughout the device and the magnetic field to be exponentially suppressed inside the \gls{sc} region with a characteristic decay length $\lambda_B$, see Supplemental Material~\cite{Supp} for more details on the model.
In the following, we employ parameters which roughly correspond to \gls{2deg} devices as the ones explored in Refs.~\cite{Hatefipour2024, MehdiHatefipour2023}.

\begin{figure}[t]
    \centering
    \includegraphics[width=\linewidth]{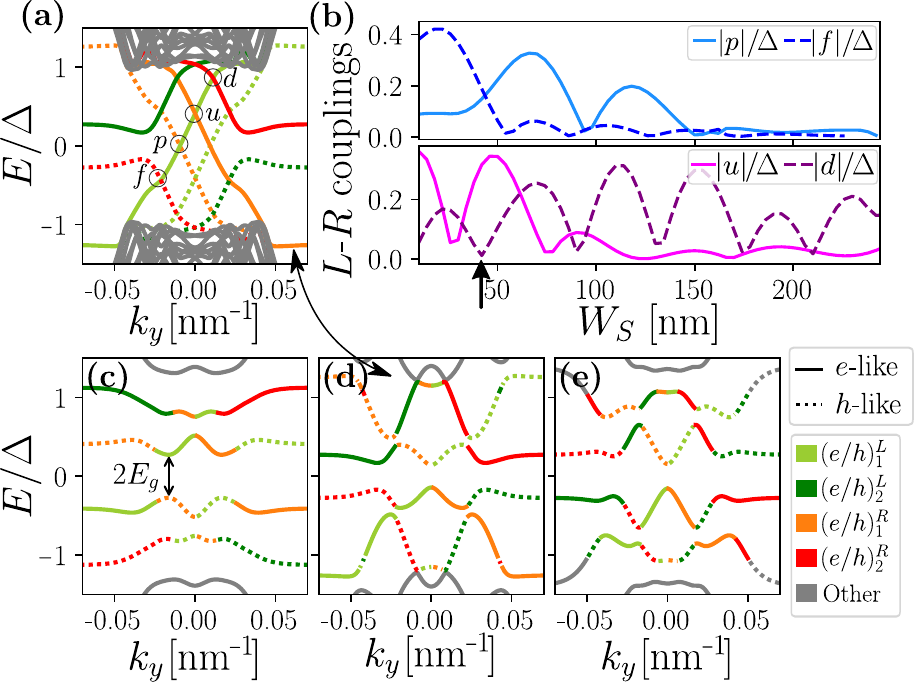}
    \caption{Band dispersion in the stripe geometry with periodic boundary conditions along $y$ for
    (a) uncoupled interfaces ($W_S\gg\xi_S$) with $\mu/\Delta = 2.1$ and (c, d, e) for $W_S = 48~\si{\nano\meter}$ with, respectively, $\mu/\Delta= 1.2, 2.1, 2.6$. 
    The colours indicate the projection over the \glspl{caes} in (a). 
    %(b) The effective couplings between \gls{caes} indicated in (a) are calculated for 
    %$\mu/\Delta= 2.1$.
    % the parameter set in (d). 
    (b) Effective couplings in (a) as a function of $W_S$ for $\mu/\Delta= 2.1$.
    }
    \label{fig:2PBC}
\end{figure}

First, we analyze the topological phases in the stripe geometry already mentioned in the introduction (periodic in the $y$ direction).
For two decoupled interfaces (i.e., in the limit $W_S \gg \xi_S$, where $\xi_S$ is the \gls{sc} coherence length), the spectrum is gapless since the \gls{l} and \gls{r} \glspl{caes} exhibit a linear dispersion inside the \gls{sc} gap $\Delta$ and do not hybridize; see \cref{fig:2PBC}(a). When the \gls{sc} region is narrowed, the \glspl{caes} form hybridized subbands with a non-zero minigap $E_g$ [\cref{fig:2PBC}(c-e)]. Depending on the width $W_S$ of the \gls{sc} stripe and on the chemical potential $\mu$, the spectrum shows different topological phases determined by the invariant $\mathcal{Q}_{\rm S}$ that signals the topological superconducting phase for $\mathcal{Q}_{\rm S}<0$. The topological invariant is defined as
\begin{align} \label{eq:pfaffian}
\mathcal{Q}_{\rm S} &= \operatorname{sign} \left\{ 
    \frac{\operatorname{Pf}\left[\mathcal{A}(k_y=0)\right]}
    {\operatorname{Pf}\left[\mathcal{A}(k_y=\pi/a)\right]}
    \right\}\, (-1)^{-\nu}~,
\end{align}
where $\nu$ is the filling factor, $a$ is the lattice constant used in the tight-binding discretization~\cite{Supp}, and $\mathcal{A}(k_y)= H(k_y) U_C$ is a skew-symmetric matrix at the time-reversal invariant momenta~\cite{ Tewari2012a, Sato2017, Heffels2023},  with $U_C$ obtained from the particle-hole symmetry $\mathcal{C} = U_C \mathcal{K}$ ($\mathcal{K}$ being the conjugation operator). 
The Pfaffian factor in~\cref{eq:pfaffian} counts the total parity of the number of crossings between different subbands at the Fermi level in half of the Brillouin zone, i.e., for $k_y\in [0, \pi/a]$. On the other hand, the factor $(-1)^{\nu}$ removes the contribution from the edge states localized at the outer edge of the stripe, not in proximity to the \gls{sc}~\cite{Shiozaki2014}, see End Matter. 
This way, $\mathcal{Q}_{\mathrm{S}}$ only takes into account the \glspl{caes} crossings, such that $\mathcal{Q}_{\mathrm{S}}<0$ indicates topological superconductivity on the stripe.

\par The coupling scheme between \gls{l} and \gls{r} states is particularly simple for filling factor $\nu=1$, since only one pair of \glspl{caes} crosses the Fermi energy. In this regime one can obtain analytical insight from an \textit{effective model} in terms of the lowest energy \glspl{caes}, see \cref{eq:MinMod} in End Matter. 
Let us denote $e^S_\nu$ ($h^S_\nu$), with $\nu=1,2$, the \gls{caes} obtained from the electron (hole) \glspl{ll} localized at the side $S=L, R$. 
In terms of these states we express the Pfaffian invariant of the effective model as 
\begin{multline} \label{eq:pfaffian_minmod}
\mathcal{Q}_{\rm S} = 
(-1)^{\nu} \,\mathrm{sign} \big[d^4 - 2 d^2 \left( E_u E_d + p^2+u^2\right)  \\
+ \left( E_d^2 + p^2-u^2 \right) \left(E_u^2 + p^2-u^2 \right) \big]~ ,
\end{multline}
where $u$, $p$, and $d$ respectively denote the coupling terms $\langle e^L_{1}|{\cal H}|e^R_1\rangle$, $\langle e^L_{1}|{\cal H}|h^R_1\rangle$, and $\langle e^L_{1}|{\cal H}|e^R_2\rangle$, indicated by the circles in \cref{fig:2PBC}(a); whereas $E_u$ and $E_d$ correspond to the energy crossings for $W_S \gg \xi_S$. 
Notice, in particular, that $p$ gives the size of the nonlocal $p$-wave coupling and $u$ is related to \gls{ec} processes connecting equal \glspl{caes} localized at the two opposite interfaces.
These coupling terms are calculated using a numerical projection on the uncoupled interface \glspl{caes}, i.e., the ones obtained in the limit of $W_S \gg \xi_S$. We represent them in \cref{fig:2PBC}(b) as a function of $W_S$. 
The couplings depend on $W_S$ according to an exponential decay given by $\xi_S$ and an oscillatory function of the Fermi wavevector $k_F$~\cite{Galambos2022, Supp}. 
The coupling term $f=\langle e^L_1|{\cal H}|h^R_2\rangle$ is neglected in~\cref{eq:pfaffian_minmod}, since it is strongly suppressed except for widths $W_S$ of the order of the localization of the \glspl{ll}, as shown in \cref{fig:2PBC}(b). 

\par The couplings' oscillatory behaviour results in an alternation of topological and trivial stripe phases as a function of the width $W_S$ and chemical potential $\mu$, according to the interplay of the different terms in \cref{eq:pfaffian_minmod}. 
Still, having a sign change in $\mathcal{Q}_{\rm S}$ at filling factor one (i.e., for $\mu < \hbar\omega/2 + V_z$, with $\omega = e B_z /m$) requires special conditions on the energy scales involved. The analysis of these conditions using the effective model gets particularly simple if we consider a case where $d=0$, like the one indicated by the arrow in \cref{fig:2PBC}(b). Then, the change of sign in $\mathcal{Q}_{\rm S}$ simply requires $E_u^2 < u^2 - p^2$, which can be understood as the competition between \gls{ec} ($u$) and \gls{car} ($p$) processes when $E_u$ is sufficiently small. This crossing energy is controlled by the interplay between the Rashba splitting $E_R = \alpha_R\sqrt{2 e B_z/\hbar}$ and the Zeeman exchange $V_z$, and we estimate the sign change to
take place for ${2}\hbar \omega/3 - V_z  \lesssim E_R \lesssim \hbar \omega$~\cite{Supp}. 

\par Going beyond the effective model, one can analyze the system phase diagram by direct diagonalization, which also allows us to
identify the main character of the subband states in the basis of the decoupled \glspl{caes}. This information is encoded in the colour code and line style of the bands plotted in \cref{fig:2PBC}(c-e), where each panel corresponds to a particular case regarding the $\mathcal{Q}_\mathrm{S}$ and $\nu$ values, in accordance with the three cases discussed in \cref{fig:1Sketch}(i-iii). 
At filling factor $\nu=1$ in the \gls{sc} topological regime ($\mathcal{Q}_\mathrm{S} = -1$) [\cref{fig:2PBC}(c)], 
the lowest energy subband is a mode coupling \gls{l} and \gls{r} states of the $\nu=1$ \gls{ll} with opposite $e$-$h$ nature, i.e., $e^L_1$ and $h^R_1$ for positive momenta. 
Since the coupled \glspl{caes} have the same $\nu$ and hence equal spin texture, this subband inversion is associated to the emergence of $p$-wave superconductivity on the stripe, in line with the invariant calculation~\cite{Clarke2014}. 

By contrast, for a topologically trivial \gls{sc} state ($\mathcal{Q}_\mathrm{S} = 1$) and $\nu=1$ [\cref{fig:2PBC}(d)], 
a band inversion at zero momentum leads to a subband coupling filled electron state $e^L_1$ and empty hole $h^R_2$ such that Andreev conversion is prohibited. 
%a band inversion at zero momentum leads to a subband coupling \gls{caes} $e^L_1$ and empty hole states $h^R_2$ such that Andreev conversion is prohibited. 
Even though $\mathcal{Q}_\mathrm{S} = 1$, the system maintains the \gls{qh} topology that manifests in a quantized \gls{ec} conductance for a finger \gls{sc}, as we show below.
Importantly, increasing the chemical potential to reach a higher filling factor $\nu=2$ yields a sign change in the topological invariant, $\mathcal{Q}_\mathrm{S}=-1$. This $\nu=2$ state now features two channels with different nontrivial topology [\cref{fig:2PBC}(e)]. 
One of the subbands is still inverted at zero momentum, with a prohibited coupling between $e^L_1$ and $h^R_1$, leading again to a perfectly quantized \gls{ec} process for the finger \gls{sc}. By contrast, the now occupied second \gls{ll} supports a nontrivial $e$-$h$ coupling that leads to the subband inversion that restores $p$-wave superconductivity for the other channel. 

\begin{figure}[t]
    \centering
    \includegraphics[width=\linewidth]{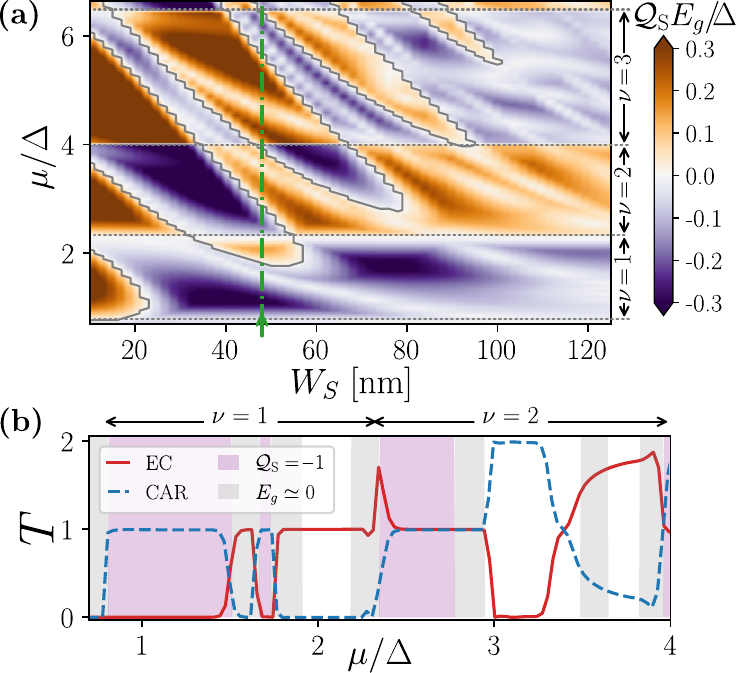}
    \caption{(a) $W_S-\mu$ map of the product of minigap and invariant, $\mathcal{Q}_{\rm S}E_g$, for a stripe geometry with parameters in~\cref{tab:paramset1}. 
    (b) Transmission probabilities for \gls{car} and \gls{ec} processes in a finger geometry of $W_S = 48~\si{nm}$ [green line in (a)] and $L_S = 500~\si{nm}$. The energy of injection is $E_\mathrm{inj}/\Delta = 10^{-3}$, see~\cite{Supp} for details on the transport calculations.
    The purple (grey) areas correspond to the $\mathcal{Q}_\mathrm{S}=-1$ (gapless) regions of a stripe geometry with the same $W_S$ and periodic boundary conditions in the $y$ direction. 
    }
    \label{fig:3GapTransp}
\end{figure}

\par The full phase diagram for the stripe geometry, calculated numerically with the set of parameters indicated in \cref{tab:paramset1} of the End Matter, is shown in \cref{fig:3GapTransp}(a). In this figure, we plot $\mathcal{Q}_\mathrm{S} E_g$ as a function of $\mu$ and $W_S$. 
Notice that not all the gapless lines are associated to topological transitions since $E_g$ also vanishes due to the oscillatory behaviour of the effective pairing between left-right \glspl{caes}. The actual topological transitions are indicated by the full grey lines, while the dashed horizontal lines indicate the changes in filling factor. As can be observed, the full grey lines define tilted tongue-like regions which widen as $\mu$ increases. These results correspond to the case of a significant \gls{rsoc} ($\alpha = 20\rm{~meV~nm}$) and a rather small field $(B=2.2\rm{~T})$. As the spin-orbit coupling is reduced the tongue-like regions move towards higher $\mu$ values, eventually above the $\nu=1$ line; whereas for higher fields these regions shrink (\cref{fig:EM:phaseMaps} in End Matter). 

\emph{Transport properties.}---
The different topological phases identified earlier have a distinct and quantized transport signal for a finite \gls{sc} finger geometry. In \cref{fig:3GapTransp}(b) the zero-energy transmission probabilities for the \gls{ec} ($T_\mathrm{EC}$, red line) and \gls{car} ($T_\mathrm{CAR}$, blue line) processes are calculated as a function of the chemical potential $\mu$ for a finite-finger configuration with fixed $W_S\times L_S$~\cite{Supp, Groth2014}. 
The gray and purple areas respectively mark the values of $\mu$ where the stripe configuration with the same $W_S$ exhibits a gapless spectrum (gray) or an invariant $\mathcal{Q}_\mathrm{S}=-1$ (purple). 

For the values of $\mu$ with a non-zero minigap in the stripe, the nonlocal conductance $G_\mathrm{nl}$ is either quantized or zero. 
Indeed, for the topologically protected states with $\mathcal{Q}_\mathrm{S}=-1$, the $\nu=1$ channel is quantized featuring perfect \gls{car} ($T_\mathrm{CAR}=1$). By contrast, the $\nu=1$ mode with $\mathcal{Q}_\mathrm{S}=+1$, where the band inversion prohibits the Andreev conversion, leads to $T_\mathrm{EC}=1$. 
Interestingly, for filling factor $\nu=2$ the two previous channels coexist when $\mathcal{Q}_\mathrm{S}=-1$, resulting in zero nonlocal conductance as the quantized \gls{ec} and \gls{car} contributions mutually cancel. 
On the other hand, for the states with $E_g \simeq 0$ in the stripe, we obtain a fragile or not even quantized $G_\mathrm{nl}$ that strongly depends on the length $L_S$ of the finger and on the energy of injection $E_\mathrm{inj}$. 
The transport results discussed in~\cref{fig:3GapTransp}(b) remain qualitatively the same for different widths $W_S$ and chemical potentials $\mu$, as shown in~\cref{fig:Sup:MapsEGM}(b) of End Matter. 

\begin{figure}[t]
    \centering
    \includegraphics[width=\linewidth]{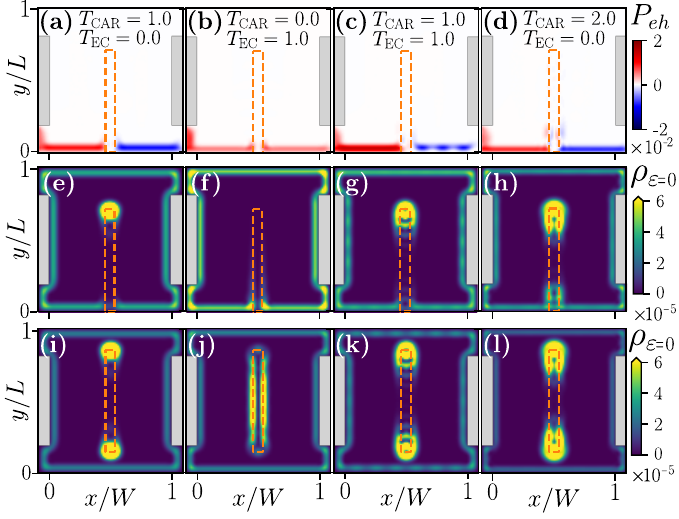}
    \caption{Scattering wavefunction and local density of states for representative cases with $W_S = 48~\si{nm}$, $L_S = 500~\si{nm}$ and total system size $W = 600~\si{nm} \times L = 700~\si{nm}$ (the injection scheme is sketched in \cref{fig:1Sketch}(a)). The columns correspond to $\mu/\Delta = \{ 1.2, 2.1, 2.6, 3.2\}$. 
    (a-d) Electron and hole content $P_{eh}(\vec{r})=|\psi_s^e(\vec{r})|^2-|\psi_s^h(\vec{r})|^2$ of the scattering wavefunction $\psi_s$ for an incoming electron at $E_\mathrm{inj}/\Delta = 10^{-3}$; 
    (e-l) local density of states at $\varepsilon = 0$ for the finger (e-h) and \gls{sc} island configurations (i-l).  
    }
    \label{fig:4ScatWfs}
\end{figure}

\par As discussed above, the quantized \gls{car} signal corresponds to the formation of pairs of Majorana modes at both ends of the finger, the lower one merging with the \gls{qh} edge channel.
% coupling to the \gls{qh} edge channel. 
To visualize the delocalization of this Majorana state into the \gls{caes} channel, in \cref{fig:4ScatWfs} we compare the electron-hole content of the scattering wavefunctions $P_{eh}$ (top panels) with the local density of states at zero energy $\rho_{\varepsilon=0}$ for the finger configuration (middle panels) and for the case of a floating \gls{sc} island detached from both sides of the \gls{qh} edges (bottom panels). 

For $\mathcal{Q}_\mathrm{S}\!=\!-1$ a pair of localized Majorana states emerge at $\nu=1$ [\cref{fig:4ScatWfs}(i)] and $\nu=2$ [\cref{fig:4ScatWfs}(k)], in correspondence with the $T_{\rm CAR}=1$ channels shown in, respectively, \cref{fig:4ScatWfs}(a) and \cref{fig:4ScatWfs}(c). The important difference being that at $\nu=2$ the \gls{car} channel coexists with another perfectly quantized \gls{ec} channel. 
When $\mathcal{Q}_\mathrm{S}=1$ we also have different behaviour depending on the filling factor. At $\nu=1$, no localized states at the tips of the \gls{sc} island are observed in \cref{fig:4ScatWfs}(j). Indeed, the scattering wavefunction [\cref{fig:4ScatWfs}(b)] and the local density of states [\cref{fig:4ScatWfs}(f,j)] feature a channel that directly connects the \glspl{ll} through the \gls{sc} region. 
By contrast, at $\nu=2$ there are two Majorana modes at the ends of the \gls{sc} [\cref{fig:4ScatWfs}(l)], leading to two quantized \gls{car} channels in \cref{fig:4ScatWfs}(d). 

\begin{figure}[t]
    \centering
    \includegraphics[width=\linewidth]{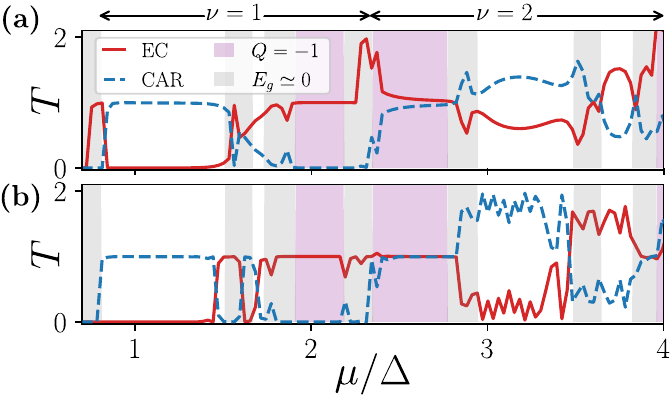}
    \caption{Transmission probabilities for the case of \cref{fig:3GapTransp}(b) with 
    (a) $E_\mathrm{inj}/\Delta = 0.05$ and 
    (b) $E_\mathrm{inj}/\Delta = 10^{-3}$ and including Anderson disorder with on-site strength $w_A/\Delta = 10$ over an extension of $50 ~\si{nm}$ around the \gls{sc} interface. }
    \label{fig:5Disorder}
\end{figure}

\par As indicated by the invariant $\mathcal{Q}_\mathrm{S}$, the Majorana states are only topologically protected when an odd number of pairs emerge separated by $L_S \gg \xi_{\rm S}$. 
For an even number, the zero-energy states form so-called \textit{fragile} Majorana pairs that can be easily gapped~\cite{Heffels2023}. 
We have verified this, as well as the robustness of the quantized transport signal to both changes in the injection energy and disorder from an on-site Anderson disorder in a region around the \gls{sc} finger.
As shown in \cref{fig:5Disorder}, the values of $\mu$ corresponding to a topological state, i.e., when $\mathcal{Q}_\mathrm{S}=-1$ for both filling factors and $T_\mathrm{EC}=1$ for $\nu=1$, are particularly stable to both perturbations. By contrast, the \gls{car} signal for two pairs of Majorana states is indeed highly sensitive to both disorder and changes in the injection energy. 
Similarly, the transmission for gapless regions is also highly sensitive to perturbations. 

\emph{Conclusions.}---
Our work reveals the existence of unconventional topological phases marked by quantized \gls{ec} nonlocal conductance in proximized \gls{2deg} devices in the \gls{qh} state. 
We have confirmed the robustness of these phases by numerical simulations using parameters representative of recent experiments~\cite{MehdiHatefipour2023, Hatefipour2024}. 
We have focused on the lowest filling factors in order to simplify the theoretical analysis and found that the topological phases displaying perfect \gls{ec} emerge at low fields and large spin-orbit splitting. 
Although reaching $\nu=1,2$ in these conditions might be experimentally demanding, the emergent phases are also present at higher filling factors [see  \cref{fig:3GapTransp}(a)] which can be within reach using state of the art techniques. 

\acknowledgments
{\it Acknowledgments.---} 
We thank R.~S\'anchez, L. Arrachea, C. Balseiro and A.~R.~Akhmerov for valuable discussions. The work was supported by Spanish CM ``Talento Program'' project No.~2019-T1/IND-14088 and No.~2023-5A/IND-28927, the Agencia Estatal de Investigaci\'on project No.~PID2020-117992GA-I00, No.~CNS2022-135950 and No.~PID2023-150224NB-I00 and through the ``María de Maeztu'' Programme for Units of Excellence in R\&D (CEX2023-001316-M). 

%%%%%%%%%%%%%%%%%%%%%%%%%%%%%%%%

%apsrev4-2.bst 2019-01-14 (MD) hand-edited version of apsrev4-1.bst
%Control: key (0)
%Control: author (8) initials jnrlst
%Control: editor formatted (1) identically to author
%Control: production of article title (0) allowed
%Control: page (0) single
%Control: year (1) truncated
%Control: production of eprint (0) enabled
%

%%%%%%%%%%%%%%%%%%%%%%%%%%%%%%%%%%%

\newpage

\subsection{End Matter}

%%%%%%%%%%%%%%%%%%

\emph{Symmetry properties and topological invariant of the stripe system.}---
The Hamiltonian $\mathcal{H}$ in \cref{eq:model:ham}, representing the proximized \gls{qh} system, belongs in class D of the Altland-Zirnbauer classification due to the intrinsic particle-hole symmetry and broken time-reversal symmetry by the external magnetic field~\cite{Schnyder2008, Ryu2010, Shiozaki2014}. 
This phase is characterized, in 1D systems, by an invariant defined by the change of sign of the Pfaffian in the particle-symmetric momenta $k_y = 0, \pi/a$, where a skew-symmetric matrix can be obtained from the full-Hamiltonian. 
We define this invariant as~\cite{ Tewari2012a, Sato2017, Heffels2023}
\begin{equation}
    \mathcal{Q}_{\rm T} = \operatorname{sign} \left\{ 
{\operatorname{Pf}\left[\mathcal{H}(k_y=0)U_C\right]}/
{\operatorname{Pf}\left[\mathcal{H}(k_y=\pi/a)U_C\right]} \right\}~,
\end{equation}
where $U_C=\tau_x\sigma_x$ is obtained from the particle-hole symmetry ${\cal C} = U_C {\cal K}$. $\mathcal{Q}_{\rm T}$ is calculated numerically using \texttt{Pfapack}~\cite{Wimmer2012}. 

\par The stripe geometry has a magnetic in-plane mirror reflection that works as an emergent time-reversal symmetry, reducing the D class to BDI. In fact, the in-plane reflection acts as
\begin{equation}
 \mathcal{A}_x \mathcal{H}(k_y) \mathcal{A}_x^{-1}= \mathcal{H}(-k_y)~, \quad \mathcal{A}_x^2 = 1~, 
\end{equation}
where $\mathcal{A}_x = \mathcal{M}_{x} \mathcal{K}$, with $\mathcal{M}_x = i P_x\tau_0\sigma_z$ and $P_x$ being the inversion operator $x \to -x$. 
Considering this emergent time-reversal symmetry, a chiral symmetry $\Gamma$ is defined by the product of time-reversal and particle-hole leading to $\Gamma = - \tau_x \sigma_y$. 
In the basis that diagonalizes $\Gamma$, we can recast the stripe Hamiltonian as a two off-diagonal blocks $D(k_y)$ with winding number $\cal W$ defined by~\cite{Chiu2016}
\begin{equation} \label{eq:sup:winding}
 \mathcal{W} = \int_{k=0}^{k=\frac{\pi}{a}} \frac{\md\theta(k)}{\pi}~,
 ~~\theta(k) = -i \operatorname{ln}\left[ \frac{\operatorname{det}[D(k)]}{|\operatorname{det}[D(k)]|}\right].
\end{equation}
The Pfaffian invariant $\mathcal{Q}_{\rm T}$ is the parity of the winding number since $\mathcal{Q}_{\rm T} = \me^{i \pi \cal W}$. 

\par
\begin{figure}[htb]
    \centering
    \includegraphics[width=\linewidth]{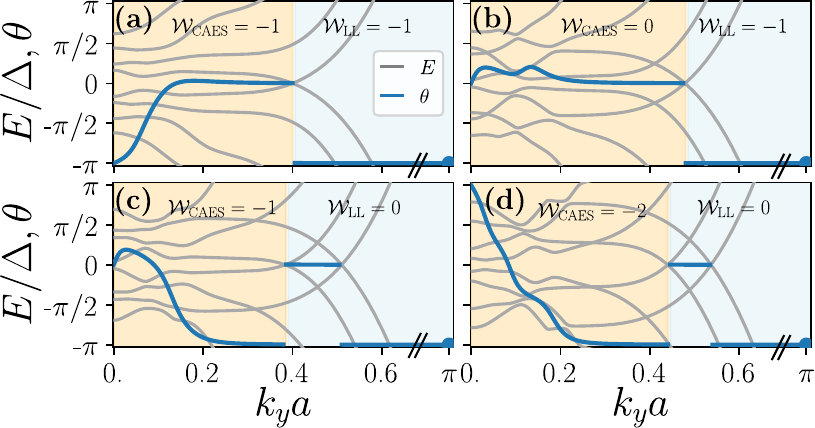}
    \caption{Winding angle $\theta$ as a function of $k_y$ in the stripe geometry. The parameters are the same as in \cref{fig:2PBC} with fixed $W_S = 48~\si{\nano\meter}$ and $\mu/\Delta= 1.2, 2.1, 2.6, 3.2$ for panels (a-d). The regions corresponding to the \glspl{caes} contributions and outer \glspl{ll} are highlighted in orange and blue, respectively. As a reference, the dispersion relation is included in grey lines. }
    \label{fig:Sup:Winding}
\end{figure}

 To separate the contributions from the \glspl{caes} and the outer \glspl{ll}, we split the integral in \cref{eq:sup:winding} as 
$\mathcal{W} = \mathcal{W}_{\rm CAES} + \mathcal{W}_{\rm LL}$, with $\mathcal{W}_{\rm CAES} =  \int_{k=0}^{k=k_{\rm LL}} \md\theta(k)/\pi$ and 
$\mathcal{W}_{\rm LL}= \int^{k=\pi/a}_{k=k_{\rm LL}} \md\theta(k)/\pi$,
where $k_{\rm LL}>0$ is the minimum (positive) momentum where the outer \gls{ll} states cross the Fermi level. 
This way, we can separate the two contributions to the Pfaffian invariant as 
\begin{equation}
    \mathcal{Q}_T = \mathcal{Q}_S \mathcal{Q}_N~,
    ~~ \mathcal{Q}_{\rm S} = \me^{i\theta(k_{\rm LL})-i\theta(0)}, 
    ~~ \mathcal{Q}_{\rm N} = (-1)^\nu ,
\end{equation}
where we have used that the Pfaffian of the outer \glspl{ll} $\mathcal{Q}_{\rm N}$ is the parity of the filling factor. 
\Cref{fig:Sup:Winding} shows the winding phase as a function of the momenta for some representative cases. 

%%%%%%%%%%%%%%%%%%%%%%%%%%%%%%%%
    \emph{Effective model for two lowest \glspl{ll}.}---
The minimal model employed to describe the coupling of the \glspl{caes} for $\nu=1,2$ states in the stripe geometry is
\begin{multline} \label{eq:MinMod}
    \mathcal{H}_\mathrm{eff} = 
    \small{\sum}_{k_y} \Psi^\dagger_{k_y} \Big[ 
    v k_y \tau_0 r_0 s_z  
    + (E_u r_u + E_d r_d) \tau_z s_0 
     \\
+ u \tau_0 r_z s_x 
+ f \tau_x r_z s_0\\
+ (p \tau_x + d \tau_0) r_y s_y 
+ M k_y^2 \tau_z r_0 s_0 
    \Big] \Psi_{k_y} ,
\end{multline} 
with
$
    \Psi_{k_y} = 
[c^{R}_{e2}, c^{L}_{e2}, c^{R}_{e1}, c^{L}_{e1}, 
 (c^{R}_{h1})^\dagger, (c^{L}_{h1})^\dagger, 
 (c^{R}_{h2})^\dagger, (c^{L}_{h2})^\dagger]^T . 
$
Here, $c^S_{t\nu}$ annihilates an electron in the \gls{caes} of side $S=L,R$ obtained from the \gls{ll} of type $t=e,h$ and index $\nu=1,2$; 
$\tau_i, r_i, s_i$, with $i = 0, x,y,z$, are the identity and Pauli matrices respectively acting on the particle-hole space, the $\nu$ index, and the interface side; and $r_{d,u} = (r_0\pm r_z)/2$. 

\par In the effective model of \cref{eq:MinMod}, the \glspl{caes} are described by a linear dispersion with velocity $v$ and with energy $E_{u(d)}$ at $k_y=0$ for the \gls{caes} with $\nu=1~(2)$. 
The coupling terms considered connect \glspl{caes} localized at opposite sides of the \gls{sc}. As defined in the manuscript, $u$ ($d$) couples states with the same particle-hole nature and same (different) index $\nu$. The coupling that mixes electron and hole states is given by $p$ ($f)$ for the same (different) $\nu$. A quadratic correction proportional to $M$ is considered to regularize the Dirac linear equation. 

\par The topological invariant in~\cref{eq:pfaffian} for the effective model is obtained analytically by evaluating the Pfaffian at $k_y=0$, resulting in 
\begin{multline}
    \mathcal{Q}_\mathrm{S}  =
    (-1)^\nu \big[ 
    d^4 +(f^2+p^2-u^2)^2 +E_u^2 E_d^2 \\
    + 2 f^2(E_u E_d +f^2)
    -(E_d^2+E_u^2) (u^2-p^2) \\
    -2 d^2 \left(E_u E_d+p^2+u^2\right) -4 d f p (E_d-E_u)\big]~.
\end{multline}
We can neglect $f$, which is strongly suppressed, to reach \cref{eq:pfaffian_minmod}. 

\par By considering the central stripe region as a superconducting Rashba-quantum well coupled to two \gls{qh} regions with well-defined \glspl{ll}~\cite{Supp} the coupling terms can be expressed as~\cite{Galambos2022}
\begin{equation} \label{eq:sup:couplings}
    c \approx c_0 \operatorname{exp}(-W_S / \xi_c) \operatorname{cos}(k_{F,c} W_S + \varphi_c)~,
\end{equation}
for $c = u, p, d, f$, such that $u_0$, $p_0$, $f_0$ and $d_0$ are real constants with units of energy and $\varphi_{c}$ takes the values $\varphi_{p} = \varphi_{f} = \pi/2$ and $\varphi_u = \varphi_d=0$. 
The Fermi momentum is approximated by $k_{F,c} = \sqrt{2m_S(\mu_S + E_c) /\hbar^2 + 1/l_R^2}$, where $E_c$ is the energy of the crossing point and $l_R = \hbar^2/(m\alpha)$. 
The decay length of the couplings inside the \gls{sc} is also considered to be energy-dependent as $\xi_{u} = \xi_{p} = \xi_S$, $\xi_d \to \infty$  and $\xi_f = \xi_S/2$, where the \gls{sc} coherence length at the Fermi level is given by $\xi_S = (\hbar/\Delta)\sqrt{2\mu/m_S}$. 

\par We fix the initial amplitudes $c_0$ 
by considering only the spin structure of the \glspl{ll}~\cite{Supp}. 
The couplings are then expressed as a function of the ratio of the Zeeman $V_z$ and the Rashba $E_R$ splitting,  
\begin{equation*}
 \beta = (\hbar\omega/2 + V_Z)\big/\sqrt{ (\hbar\omega/2 + V_Z)^2+E_R^2}~,
\end{equation*}
as
$u_0 \approx \Delta \beta$, 
$p_0 \approx \Delta\sqrt{1-\beta^2}$, 
$d_0 \approx \Delta\sqrt{(1-\beta)/2}$,  
$f_0 \approx \Delta \sqrt{(1+\beta)/2}$. 
In the absence of \gls{rsoc}, $p_0=d_0=0$.

\par
\begin{figure}[htb]
    \includegraphics[width=\linewidth]{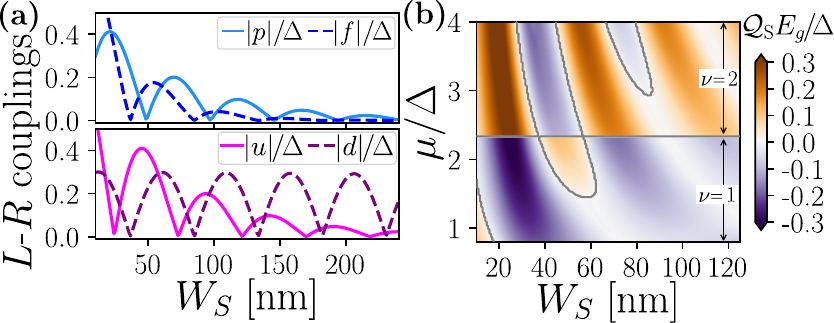}
    \caption{Effective model. 
    (a) Dependence of the coupling terms on $W_S$ at fixed $\mu/\Delta = 2.1$. (b) Phase diagram for the region of $\nu=1$ and $\nu=2$. 
    We estimate $E_u/\Delta = 0.35$, $E_d/\Delta = 1.1$, with the other parameters from \cref{tab:paramset1}.
}
    \label{fig:Sup:minmod}
\end{figure}

 Using these approximations, the $W_S$ dependency of the coupling terms is computed in \cref{fig:Sup:minmod}(a) for a fixed $\mu$. The estimated values can be compared to the numerically computed projectors in \cref{fig:2PBC}(a). 
The phase diagram of the effective model is calculated in \cref{fig:Sup:minmod}(b), showing a qualitative agreement with the complete calculation in~\cref{fig:3GapTransp}(a).

\emph{Comments on parameters.}---
The parameter values used in the plots of the main text are listed in~\cref{tab:paramset1}. They have been chosen to be representative of $\rm Ga_xIn_{1-x}As/Nb$ devices where the effective mass is estimated of the order of $m^* \approx 0.02 -0.07 m_e$ \cite{Adachi2009}, the effective $g$-factor is $|g^*|\approx 10-15$ \cite{Yuan2020} and $\alpha_R \gtrsim 10~\si{\milli \eV \nm}$~\cite{Lotfizadeh2024} or even higher~\cite{Wickramasinghe2018, Farzaneh2024, Escribano2022}. 

\begin{table}[htb] 
    \begin{tabular}{|c|}
        \hline
    $ \Delta = 1~\si{\milli \electronvolt},$ ~~ $a = 2~\si{\nano \meter},$  
    ~~ $m_\mathrm{eff}/m_e \simeq 0.07,$ 
    ~~ $B_z = 2.2~\si{\tesla},$\\
    $\alpha_R = 20~\si{\milli\eV\nano \meter}~,$ ~~ $g = -8~,$ ~~ $\lambda_B = 6~\si{\nano\meter}~.$ \\
    \hline
    \end{tabular}
    \caption{Parameters for the main text calculations.}
    \label{tab:paramset1}
\end{table}    

\par
\begin{figure}[htb]
    \centering
    \includegraphics[width=\linewidth]{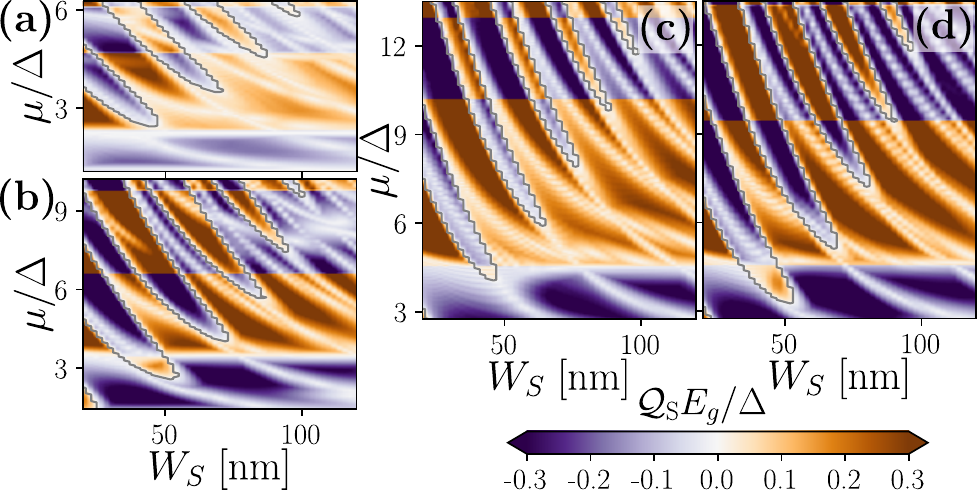}
    \caption{
    Phase diagrams varying $\alpha_R$ and $B_z$. For panels (a-d) the \gls{rsoc} and the magnetic field are respectively set to $\alpha_R ~\si{[meV~nm]}= 10, 20, 10, 20$ and $B_z~{\rm [T]}  = 2.2, 3.5, 5, 5$. 
    }
    \label{fig:EM:phaseMaps}
\end{figure}

 Phase diagrams for other sets of parameters are shown in~\cref{fig:EM:phaseMaps}. The resulting topological regions exhibit different extensions and appear at different values of $\mu$ and $W_S$ depending on the \gls{rsoc} and magnetic field when compared to \cref{fig:3GapTransp}(a). 
As can be observed in~\cref{fig:EM:phaseMaps}(a), using the same field as~\cref{fig:3GapTransp}(a) but with smaller $\alpha_R$, the lowest tongue-like region does not reach filling factor $\nu=1$. At higher fields and lower \gls{rsoc} [e.g.,~\cref{fig:EM:phaseMaps}(c)] the tongue-like regions get narrower. The variation of the maps with other parameters is illustrated in Ref.~\cite{Supp}.

\begin{figure}[htb]
    \centering
    \includegraphics[width=\linewidth]{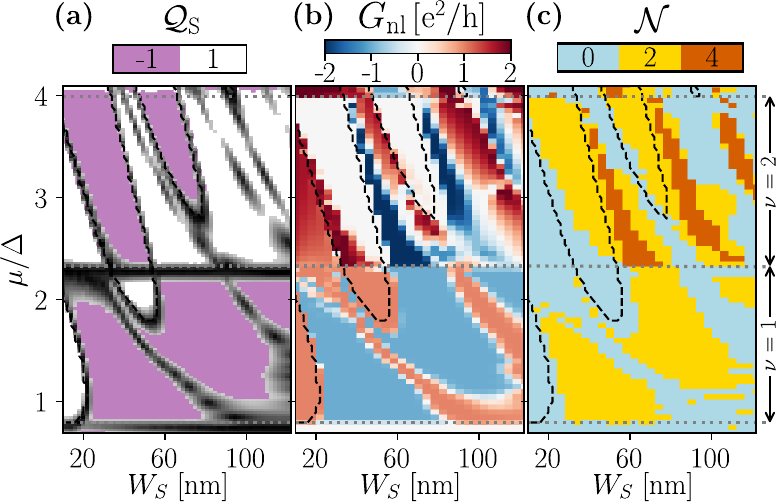}
    \caption{(a) Phase diagram showing the Pfaffian invariant $\mathcal{Q}_\mathrm{S}$ and the gapless regions (grey) for a stripe configuration. (b) Conductance $G_\mathrm{nl}$ in a finger geometry. (c) Number of Majorana zero modes $\mathcal{N}$ in a \gls{sc} island configuration. The parameters used are listed in \cref{tab:paramset1}. 
    }
    \label{fig:Sup:MapsEGM}
\end{figure}

\emph{Consistency between phase diagram, nonlocal conductance and Majorana number.}---
\Cref{fig:Sup:MapsEGM} illustrates the consistency between the $\mathcal{Q}_{\rm S}$ phase diagram for the stripe configuration (a) with the $G_{\rm nl}$ for the finger geometry (b), and the number $\mathcal{N}$ of Majorana zero modes in the \gls{sc} island (c).

\clearpage

\onecolumngrid

\setcounter{equation}{0}
\renewcommand{\theequation}{S\arabic{equation}}
\crefname{equation}{Eq.}{Eqs.}
\Crefname{equation}{Equation}{Equations}

\setcounter{figure}{0}
\renewcommand{\thefigure}{S\,\arabic{figure}}
\crefname{figure}{Fig.}{Figs.}
\Crefname{figure}{Figure}{Figures}

\section{Supplemental Material to ``Emergent topology by Landau level mixing in quantum Hall-superconductor nanostructures''}

In this supplemental material we provide further details on our model for a proximized \gls{2deg} in the \gls{qh} regime, including the spatial dependence assumed for all parameters and the discretization procedure used in the numerical calculations. We also describe the approximations considered for an analytical description of the phase diagrams within an effective model and discuss the finite-size effects on the transport porperties in the finger and island geometries. Finally, we compute numerical phase diagrams for other sets of parameters than those considered in the main text.  
 
\section{Details on the model and numerical calculations} \label{sec:Sup:Model}
\emph{Model Hamiltonian}.---
As stated in the main text, the general Hamiltonian that we consider can be written as
\begin{align} \label{eq:app:ham}
    \mathcal{H} = \left( \frac{\hbar^ 2 \vec{k}^ 2}{2 m}
    -\mu(\vec{r}) \right) \sid \tz 
    + V_z (\vec{r}) \sz \tid
    + \alpha_R(\vec{r}) \left(k_x \sy \tid - k_y \sx \tid \right) - \Delta (\vec{r}) \tx \sz~,
\end{align}
where $\vec{k}=-i\vec{\nabla}_r$, while $\sigma_i$ and $\tau_i$ are Pauli matrices and correspond to the spin and electron-hole space, respectively. 
All potentials defined in \cref{eq:app:ham} are space dependent. 
In this way, the proximized \gls{sc} region can be conveniently defined by space-dependent functions.
In the results presented in the main text, the \gls{rsoc} $\alpha_R(\vec{r})$ is assumed to be constant all over the device and the Zeeman splitting $V_z$ is assumed to decay in a Gaussian-like shape within the \gls{sc} region as
\begin{subequations}
\begin{align} \label{eq:mod:aR_VZ}
    V_z (\vec{r})= \begin{cases}
        V^N_z, \quad &~\text{if~}\vec{r}\in\text{N~region},\\
        V^N_z \exp\left(-\frac{|\vec{r-r_\mathrm{NS}}|^2}{2\lambda_z^2}
        \right)~, &~\text{if~}\vec{r}\in\text{SC~region},
        \end{cases}
    % f(\vec{r})  = \begin{cases}
    %     f_0, \quad ~\text{if~}\vec{r}\in~\text{N~region}~,\\
    %     f_0 \exp(-\frac{|\vec{r-r_\mathrm{NS}}|}{d_f})~, ~\text{if~}\vec{r}\in~\text{SC~region}~.
    % \end{cases}
\end{align}
\end{subequations}
% where $f = \{ \alpha_R, V_Z\}$ and $d_f = \{d_R, d_Z\}$. 
where the vector $\vec{r}_\text{NS}=(x_{\rm NS}, y_{\rm NS})$ indicates the position of the nearest N-\gls{sc} interface. The Zeeman splitting is given by the magnetic field as $V^N_z = g^* \mu_B B_z/2$, where $\mu_B$ is the Bohr magneton and $g^*$ is the effective $g$ factor. 
The magnetic field is included via the vector potential, for which we choose the following gauge
\begin{align}
    \vec{A}(\vec{r}) & = (0, A_y(x), 0)~, \quad \text{with~}
    A_y(x)  = 
    \begin{cases}
        B_z(x-x_{\mathrm{NS}}+\zeta_x\lambda_B),
        \quad &\text{if } x \in \text{N region},\\
         \zeta_x B_z \lambda_B \exp\left(-\frac{|x-x_{\mathrm{NS}}|}{\lambda_B} \right), 
         \quad &\text{if } x \in \text{SC region},
    \end{cases} \nonumber
\end{align}
where $\zeta_x = \operatorname{sign}(x-x_{\mathrm{NS}})$. In the finger geometry, the regions of fixed $y$ without a \gls{sc} interface are defined by the aforementioned $A_y(x)$ with $\zeta_x = 0$. 
The chemical potential is defined by
\begin{equation}
   \mu(\vec{r}) = 
    \begin{cases}
        \mu_N, \quad &\text{if } \vec{r} \in \text{N region},\\
         \mu_S, \quad &\text{if } \vec{r} \in \text{SC region}.
    \end{cases} 
\end{equation}
If not stated otherwise, we consider $\mu \equiv \mu_N = \mu_S$.
Finally, the $s$-wave pairing is considered nonzero only in the region proximitized by the \gls{sc}:
\begin{equation}
   \Delta(\vec{r}) = 
    \begin{cases}
        0,
        \quad &\text{if } \vec{r} \in \text{N region},\\
         \Delta, \quad &\text{if } \vec{r} \in \text{SC region}.
    \end{cases} 
\end{equation}
% The correspondent magnetic length is $l_B = \sqrt{\hbar /(|e|B_z)}$, or, equivalently, as a function of the magnetic flux 
% $$l_B = a \sqrt{\frac{1}{2\pi}\frac{\Phi_0}{\Phi}}~,$$ 
% with $\Phi = B_z a^2$ and $\Phi_0 = h/|e|$. 

\par \emph{Tight-binding description}.---
The Hamiltonian~\eqref{eq:app:ham} is discretized in a square lattice with lattice parameter $a$ as:
\begin{multline}
    \mathcal{H}_\mathrm{2D} = 
        \sum_{i,j} \Psi^{\dagger}_{i,j}
            \Bigl( \left[ 4t -\mu_{i,j} \right] \sid \tz - \Delta_{i,j} \sz \tx
                   + V^z_{i,j} \sz \tid \Bigr)~ \Psi_{i,j} 
         + \sum_{\langle \text{NN-x}\rangle} \Psi^{\dagger}_{i,j} 
         % e^{-i\phi_1 (y) \tz} 
         \left(
            -t \sid \tz -i \frac{\alpha_{i,j}}{2a} \sy \tid  \right) \Psi_{i+1,j} + \text{H.c.}~\nonumber \\
        + \sum_{\langle \text{NN-y}\rangle} \Psi^{\dagger}_{i,j} e^{-i \tz \phi_{i,j}} \left(
            -t \sid \tz + \frac{i \alpha_{i,j}}{ 2 a} \sx \tid  \right) \Psi_{i,j+1} + \text{H.c.}~,
\end{multline}
where $(i,j)$ indicate the lattice-site indices. The sums in the previous expression are performed over \gls{nn} sites in the $x$ and $y$ directions. The hopping term is $t = {\hbar^ 2 }/{(2 m a)}$ and the basis used is $\Psi_{i, k_y} = ( c_{i,j,\up}~c_{i,j,\dw}~c^\dagger_{i,j,\dw}~c^\dagger_{i,j,\up})$.

The vector potential is included through the Peierls phase by replacing the hopping between \gls{nn} in the $y$ direction by
$t_{y} \rightarrow t_{y} \exp\left( - i \tau_z \phi_{i,j} \right)$, with the phase
\begin{align}
    \phi_{i,j} & \equiv \frac{e}{\hbar} (\vec{r}_{i,j+1}-\vec{r}_{i,j})\frac{\vec{A}(\vec{r}_{i,j+1}) + \vec{A}(\vec{r}_{i,j})}{2} ~.
    \label{eq:App:Magn_phi}
\end{align}
Here, $\vec{r}_{i,j}$ indicates the position of the site with lattice indices $(i,j)$. The terms $f_{i,j}$, with $f = \{\mu, \Delta , V_z, \alpha \}$, are defined from the continuum functions of the potentials as $f_{i,j} \equiv f (\vec{r}_{i,j}$). 

\par The Anderson disorder included in \cref{fig:5Disorder}(b) is introduced by defining the chemical potential as
\begin{equation}
   \mu(\vec{r}) = 
    \begin{cases}
        \mu_N + w_A \eta(\vec{r})~, \quad &\text{if~}|\vec{r}-\vec{r}_\mathrm{NS}|<d_A~,\\
         \mu_S + w_A \eta(\vec{r})~,
         \quad &\text{if } \vec{r} \in \text{SC region}, \\
         \mu_N~,
         \quad &\mathrm{otherwise}, 
    \end{cases} 
\end{equation}
where $\eta(\vec{r})\in [-0.5, 0.5)$ is a uniformly-distributed random value and $d_A$ defines a region near the N-\gls{sc} interface where a disordered interface is considered.

\par \emph{Details on the numerical calculations and the transport setup}.---
The transport calculations are performed using \texttt{Kwant}~\cite{Groth2014} in a setup as the one depicted schematically in \cref{fig:1Sketch}(a). The leads considered are semi-infinite in the $x$ direction and finite in $y$ direction with a width of $W_\mathrm{leads }= 140~\si{nm}$. They are described by \cref{eq:app:ham} with $\alpha_R = \Delta = V_z=0$, the chemical potential is fixed at $\mu_L/\Delta = 4.4$. 

\texttt{Kwant}~\cite{Groth2014} is employed to extract the scattering matrix $\mathcal{S}$ for a two-lead system as the one sketched in \cref{fig:1Sketch}(a). The scattering matrix elements are $\mathcal{S}^{e,t}_{\alpha, \beta}$, with $\alpha$ being an electronic ($e$) incoming mode at the $in$ lead and $\beta$ being an outgoing mode of type $t= e,h$ in the $out$ lead. 
% $\mathcal{S}$ is obtained for the set of $N^i_e$ incoming electronic modes to the $N^o_e$ ($N^o_h$) outcoming electron (hole) type propagating states of the lead. 
The transmission probability is then obtained by the Landauer-Büttiker formalism at zero-temperature as:
\begin{align}
    T_{\rm CAR} (E_{\rm inj}) & = 
    \sum_{\alpha, \beta}
    \left| \mathcal{S}^{e,h}_{\alpha \beta}(E_{\rm inj})\right|^2~,\\
    T_{\rm EC} (E_{\rm inj}) & = 
    \sum_{\alpha, \beta}
    \left| \mathcal{S}^{e,e}_{\alpha \beta}(E_{\rm inj})\right|^2~,
\end{align}
such that the nonlocal conductance is given by
\begin{equation}
    G_{\rm nl} (E_{\rm inj})= \frac{e^2}{h}\left[T_{\rm EC}(E_{\rm inj}) - T_{\rm CAR}(E_{\rm inj})\right]~.
\end{equation}
In the transmission probability results shown in \cref{fig:2PBC}(b) and \cref{fig:5Disorder}, the numerical criteria for defining the $E_g \simeq 0$ is set to $E_g/\Delta \leq 0.05$. 

\par The stripe geometry is considered infinite in the $y$ direction and of width $W = 400~\si{nm}$ in the $x$ direction.  
In the finger geometry, the calculations in \cref{fig:3GapTransp}(b) are performed for a system of size $W \times L = 400~\si{nm} \times 600~\si{nm}$. 
For the \gls{sc} island configuration in \cref{fig:4ScatWfs}, the system considered has extension $W \times L = 600~\si{nm} \times 700~\si{nm}$. 

\par Together with the scattering matrix, the wave function formulation employed in \texttt{Kwant}~\cite{Groth2014} returns as main output the scattering wavefunction $\psi_{s, \alpha}(\vec{r})$ that solves the propagation problem of an incoming mode $\alpha$ from lead $in$ in the scattering region.
In the manuscript, the observables calculated over the scattering wavefunction are defined as
\begin{equation*}
    \langle \psi_s(\vec{r})
    \rangle_{\widehat{\mathcal{O}}} =       \sum_{\alpha} \langle 
    \psi_{s, \alpha} (\vec{r})
    \rangle_{\widehat{\mathcal{O}}}~,
\end{equation*}
where $\widehat{\mathcal{O}}$ denotes the observable operator. 
In particular, the electron and hole content of the scattering wavefunction represented in \cref{fig:4ScatWfs} is defined as
\begin{equation}
P_{eh}(\vec{r}) = 
|\psi_s^e(\vec{r})|^2-|\psi_s^h(\vec{r})|^2~,
\quad |\psi_s^t(\vec{r})|^2 = \sum_\alpha |\langle \psi_{\alpha,s}(\vec{r}) \rangle_{\hat{Q}_t}|^2~,
\end{equation}
with $t = e,h$ and where $\hat{Q}_{e(h)}$ is the projector over the electron (hole) Nambu space. 

The local density of states $\rho(\varepsilon)$ represented in the second and third rows of \cref{fig:4ScatWfs} is defined as:
\begin{equation}
    \rho(\varepsilon, \vec{r}) 
    = \operatorname{Im} \left[ \lim_{\eta \to 0^+} \sum_m \frac{|\psi_m(\vec{r})|^2}{\varepsilon-E_m+i\eta} \right]~,
\end{equation}
where $\psi_m(\vec{r})$ is the $m$-th eigenstate of the system with energy $E_m$ obtained by sparse numerical diagonalization. 

% \yuriko{Remove? :
% The scattering wavefunction can be related to the Green function of the scattering region $\mathcal{G}_S$ as
% \begin{equation}
%     \psi_{s, \alpha} = \mathcal{G}_s Q_\alpha^-~,
% \end{equation}
% %
% where $Q_\alpha^-$ is a projector into the mode $\alpha$, see Ref.~\cite{Santos2019} for more details. 
% }

\section{Effective analytical model}

\emph{Landau levels with \gls{rsoc}}.---
In the limit of $\lambda_B = 0$ and $\lambda_z=0$, the bulk electronic \glspl{ll} with \gls{rsoc} are defined by the wavefunctions $\Phi^S_{e, (n,\chi)}$ with energy $E_{(n,\chi)}$, where
\begin{subequations} \label{eq:app:lls}
\begin{align}  
    \Phi^S_{e, (n,\chi)} = & \sqrt{(1+ \chi\beta_n)/2} \ket{n, S,\up} + \operatorname{sign(S)} \sqrt{(1- \chi \beta_n)/2}\ket{n-1, S, \dw} ~, \label{eq:app:ll:wfs}\\
    E_{(n, \chi)} = & \hbar \omega n + \chi \sqrt{
    \left( \hbar \omega/2+ V_z \right)^2 + n E_R^2  
    }~,
\end{align} \end{subequations}
with 
$ \beta_n = ( \hbar \omega/2+ V_z )/{\sqrt{ \left( \hbar \omega/2+ V_z \right)^2+nE_R^2}}~$ and 
$E_R = \alpha_R\sqrt{2 e B_z/\hbar}$~\cite{Shen2004, Schliemann2003}. 
In these expressions, $\chi = \pm 1$ for $n\geq 1$ and $\chi=1 $ for $n=0$. The eigenstates $\ket{n, S, \up(\dw)}$ represent the spin-polarized $n$-th \glspl{ll} with spin direction $\up (\dw)$ and localized at the side $S = L,R$. 
In the following, we use the abbreviated notation in which the lowest-energy state is labelled as $\nu=1$ and corresponds to the state with $(n,\chi)=(1,-1)$. The state labelled as $\nu=2$ is given by $(n,\chi)=(0,1)$. 
To simplify the notation we also use $\beta \equiv \beta_1$. 

% \begin{equation} \label{eq:SplttingOmega}
%     \Omega = - \frac{\hbar\omega}{2} - V_z + \frac{1}{2} \sqrt{\left( \hbar \omega -{2 V_z}\right)^2 + {E_R}^2}
% \end{equation}

\par \emph{Coupling terms}.---
If we neglect the spatial localization of the \glspl{ll} in the perpendicular direction to the interface, we can consider the superconducting gap as mainly coupling the spin structure of the \gls{ll} states. This way, the $\rm L$-$\rm R$ couplings of the minimal model can be estimated by the following overlap terms: 
\begin{subequations} \label{eq:app:couplings0} 
\begin{align}
    u_0 & \sim |\braket{\Phi_{e1}^L|\Delta\tau_0\sigma_0}{\Phi_{e2}^R}| \approx \Delta \beta~,\\
    %{\Delta E_{\nu2}}/{\sqrt{E_{\nu2}^2+E_R^2}}~,\\
    p_0 & \sim |\braket{\Phi_{e1}^L|\Delta\tau_x\sigma_z}{\Phi_{h1}^R}| \approx \Delta\sqrt{1-\beta^2}~ ,\\
    % {\Delta E_R }/{\sqrt{E_{\nu2}^2+E_R^2}}~ ,\\
    d_0 & \sim |\braket{\Phi_{e1}^L|\Delta\tau_0\sigma_0}{\Phi_{e2}^R}| \approx \Delta\sqrt{(1-\beta)/2}~, \\
    % \frac{\Delta}{\sqrt{2}} \left( {1 - \frac{E_{\nu2}}{\sqrt{E_{\nu2}^2+E_R^2}}} \right)^{1/2}~, \\
    f_0 & \sim |\braket{\Phi_{e1}^L|\Delta\tau_x\sigma_z}{\Phi_{h2}^R}| \approx \Delta \sqrt{(1+\beta)/2}~,
    % \frac{\Delta}{\sqrt{2}} \left( {1 + \frac{E_{\nu2}}{\sqrt{E_{\nu2}^2+E_R^2}}} \right)^{1/2}~,    
\end{align}  \end{subequations}
where $\Phi_{t\nu}^S$ are the bulk \glspl{ll} localized at side $S = L,R$, of particle type $t = e,h$ and filling factor $\nu = 1,2$ given by~\cref{eq:app:ll:wfs}. Note that the approximate values for the couplings depend only on the factor $\beta$ that accounts for the relation of the Zeeman and the Rashba splitting in the bulk \glspl{ll}. 

\par \emph{Conditions for band inversion}.---
To derive the approximate conditions for $\mathcal{Q}_{\rm S}=-1$ presented in the manuscript, we use the estimated values for the coupling constants from Eqs.~\eqref{eq:app:couplings0} in the analytical Pfaffian invariant of \cref{eq:pfaffian_minmod}. 
In the simplest case of neglecting $d$, we obtain
\begin{equation}
    \mathcal{Q}_{\rm S} = (-1)^\nu \operatorname{sign}\Big\{
     \left[E_d^2 +\widetilde{\Delta} ^2 (\sin^2 (k_F W_S)-\beta^2) \right] 
     \left[E_u^2 + \widetilde{\Delta} ^2 (\sin^2 (k_F W_S)-\beta^2) \right]\Big\}~,
\end{equation}
where $\widetilde{\Delta} = \Delta e^{-W_S/\xi_S}$.
Since $E_d \gtrsim \Delta$ and $0 < \beta \leq 1$, the sign change is only given by the second term and, in particular, its minimal value is obtained for $k_F W_S = n \pi $ with $n = 1,2,...$, i.e., when $p= 0$. 
This condition fixes the width, such that, by using $k_F \approx \sqrt{2m \mu}/\hbar$ and $\xi_S = (\hbar/\Delta)\sqrt{2\mu/m}$, we obtain
\begin{equation} \label{eq:app:approxQs}
    \lim_{p, d\to 0}\mathcal{Q}_{\rm S} \approx 
    (-1)^\nu \operatorname{sign} \left[E_u^2 - \beta^2 \Delta^2 e^{-n\pi \Delta/\mu} \right]~,
\end{equation}
where, in the following, we only consider the $n=1$ case. 

\par We can estimate $E_u$ from the subgap expression for the \glspl{caes} dispersion from Refs.~\cite{Hoppe2000, Giazotto2005}, considering the \gls{rsoc} as a constant shifting of the bulk bands, namely, 
\begin{equation} \label{eq:app:Eu}
    E_u \approx 2\Delta\frac{\hbar \omega 
    \pm \Gamma_0
    -\sqrt{\left(
    \frac{\hbar \omega}{2} + \frac{V_z}{2}\right)^2 + \frac{E_R^2}{2}}
    }{\hbar \omega/\pi + \Delta}~,
\end{equation}
where $\Gamma_0 = \frac{\hbar \omega }{2\pi} \arccos(\Omega_0)$, with $\Omega_0\to 0$ for an ideal N-\gls{qh} interface. 
Assuming that $\hbar \omega \gg E_R> V_z$, we introduce \cref{eq:app:Eu} in \cref{eq:app:approxQs} and expand it up to second order in $E_R$ around the $\mu$-independent solution $E_R^0 = \sqrt{ 5/8 (\hbar\omega)^2 -V_z(V_z/2 +  \hbar\omega)}$. As a result, the condition for $\mathcal{Q}_{\rm S} = -1$ in \cref{eq:app:approxQs} can be approximated as
%Employing this expression, we make a series expansion, up to second order, of \cref{eq:app:approxQs} for $E_R$ around the $\mu$-independent solution $E_R^0 = \sqrt{ 5/8 (\hbar\omega)^2 -V_z(V_z/2 +  \hbar\omega)}$ and for $\hbar \omega \gg E_R> V_z$. In this regime, the condition for $\mathcal{Q}_{\rm S} = -1$ in \cref{eq:app:approxQs} can be approximated by:
%
\begin{equation}
    2\hbar \omega /3- V_z  < E_R <
   \hbar \omega - V_z/3~,
\end{equation}
which is further simplified in the text as $2\hbar/3 \omega - V_z  < E_R \lesssim \hbar \omega$. 

\par In the more general case of considering $d \neq 0$, the condition of band inversion when $p=0$, i.e., for $k_F W_S = n \pi $, is
\begin{equation}
    \lim_{p \to 0}\mathcal{Q}_S = (-1)^\nu\big\{ \widetilde{\Delta}^4 \beta^4
    - \widetilde{\Delta}^2 \beta^2 \left[ E_u^2+E_d^2 - \Delta^2 (1-\beta)  \right]
    + \left[ E_uE_d-(1-\beta) \Delta/2 \right]^2\big\}~,
\end{equation}    
that leads to a range of $E_u$ given by $E_{lim,-} < E_u < E_{lim,+}$, where
\begin{equation} \label{eq:app:Elim}
    E_{lim, \pm } = \frac{(1-\beta) \Delta^2/2 
    - \beta ^2 \widetilde{\Delta}^2  
    \mp \beta \widetilde{\Delta} E_d }
    {E_d \pm \beta  \widetilde{\Delta}}~.
\end{equation}
Without further assumptions on $E_u$, \cref{eq:app:Elim} allows us to estimate the values of $E_u$ needed to obtain a band-inversion regime for fixed values of \gls{rsoc} and Zeeman splitting. 

\par On the other hand, an estimation of the \gls{rsoc} required to reach $\mathcal{Q}_{\rm S}=-1$ is obtained in the $d\neq 0$ regime by considering $E_d \approx \Delta$, $E_u$ as in \cref{eq:app:Eu}, and $\hbar \omega \gg E_R> V_z$, leading to the approximate condition
\begin{equation}
    \frac{\Delta}{3} + \sqrt{\frac{\hbar\omega}{3}(V_z+\Delta)}
   \lesssim  E_R -\frac{\hbar\omega}{2} 
   \lesssim \frac{\hbar\omega}{3}  - \frac{3\Delta^2}{2\hbar\omega}~.
\end{equation}

\par \emph{Symmetries of the effective model}.---
The minimal model in~\cref{eq:MinMod} is obtained by imposing the following symmetries:
\begin{itemize}
\item Particle-hole symmetry, represented in the basis of the two lowest \gls{caes} as $\Tilde{\mathcal{C}} = \tau_x r_x s_0 \mathcal{K}$ so that
\begin{equation}
    \Tilde{\mathcal{C}} 
    \mathcal{H}_{\mathrm{eff}} (k_y)\Tilde{\mathcal{C}}^{-1} = 
    -\mathcal{H}_{\mathrm{eff}} (-k_y)~.
\end{equation}
\item Magnetic in-plane mirror reflection that relates the Hamiltonian for the L and R sectors in the case of the stripe geometry. 
In the basis of the minimal model, we represent it as $ \Tilde{M} = \tau_0 r_z s_x$ and require that
\begin{equation}
    \Tilde{M} \mathcal{H}_{\mathrm{eff}}(k_y) \Tilde{M}^{-1} = 
    \mathcal{H}_{\mathrm{eff}} (-k_y)~.
\end{equation}
\item Spin polarization in $S_z$ in the limit of $\alpha_R\to 0$. In the absence of \gls{rsoc}, the \glspl{caes} are spin-polarized and the spin operator in $z$ commutes with the Hamiltonian, i.e., $[\mathcal{H}_\mathrm{eff}, S_z] = 0$, where $S_z = \tau_0 r_z s_0$. 
This is indeed fulfilled by the effective model for $p \to 0$ and $d \to 0$, terms that are only non-zero when including \gls{rsoc}. 
\end{itemize}

\section{Finite-size effects}

\par
\begin{figure}[htb]
    \centering
    \includegraphics[width=0.45\linewidth]{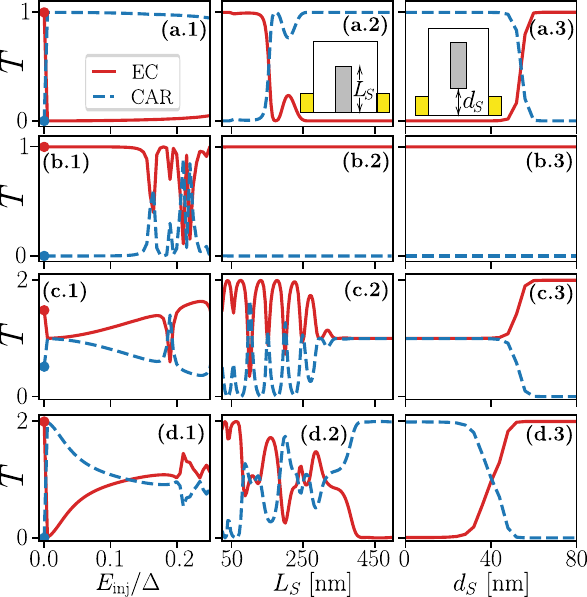}
    \caption{Transmission probabilities for the cases of \cref{fig:4ScatWfs} as a function of $E_\mathrm{inj}$ (column 1), the length of the finger $L_S$ as sketched in the inset of (a.2)(column 2), and the distance $d_S$ to the \gls{qh} edge in a \gls{sc} island geometry (column 3), see inset in (a.3). The rows, denoted by (a-d), correspond to $\mu/\Delta =  1.2, 2.1, 2.6, 3.2$, respectively. $W_S = 48~\si{nm}$ in all plots, in columns 2 and 3 we set $E_{\rm inj}/\Delta = 10^{-3}$, and fix $L_S = 500~\si{nm}$ in columns 1 and 3. }
    \phantomsection
    \label{fig:Sup:MajoDep}
\end{figure}

 In this section we briefly discuss finite-size effects on the transport properties for the finger and island configuration, as summarized in \cref{fig:Sup:MajoDep}. 
Due to these effects, the transmission probabilities at $E_{\rm inj} = 0$ do not necessarily reflect the expected quantized values for each topological phase. This can be seen in the first column of \cref{fig:Sup:MajoDep} where the energy dependence of the transmission probabilities is represented for the four representative cases discussed in \cref{fig:4ScatWfs}, i.e., for devices with parameters corresponding to a stripe with $\mathcal{Q}_{\rm S} = \pm 1$ and $\nu=1,2$. While the quantized \gls{ec} in $\nu=1$ displays no abrupt change at low injection energies [\cref{fig:Sup:MajoDep}(b.1)], the transmission for $E_\mathrm{inj}/\Delta<10^{-3}$ differs significantly for the other cases involving the appearance of one [\cref{fig:Sup:MajoDep}(a.1) and \cref{fig:Sup:MajoDep}(c.1)] or two [\cref{fig:Sup:MajoDep}(d.1)] \gls{car} channels. 
Regarding the energy-dependence at higher values, the $\nu=1$ states are quite more resilient and show perfectly quantized signals up to $E_{\rm inj} \lesssim 0.2\Delta$. 

The second column in \cref{fig:Sup:MajoDep} studies the evolution of $T_{\rm CAR}$ and $T_{\rm EC}$ as a function of the length $L_S$ of the \gls{sc} finger. As expected, the pair of Majorana states formed at each ends of the finger have to be spatially separated to prevent their hybridization, see \cref{fig:Sup:MajoDep}(a.2), \cref{fig:Sup:MajoDep}(c.2) and \cref{fig:Sup:MajoDep}(d.2), and a quantized signal is obtained for $L_S \gtrsim 4 \xi_S$. 
Finally, in column 3, the \gls{sc} island configuration is studied as a function of the distance $d_S$ defined between the lower end of the \gls{sc} region and the \gls{qh} edge, see inset in \cref{fig:Sup:MajoDep}(a.3). In the limit of $d_S\gg l_B$, the island configuration with \gls{sc} topological character develops localized Majorana states at the two ends of the \gls{sc} island, see \cref{fig:4ScatWfs} panels (i,k,l). By decreasing $d_S$, the Majorana state localized at the lower edge couples with the \gls{ll}. As shown in panels \cref{fig:Sup:MajoDep}(a.3), \cref{fig:Sup:MajoDep}(c.3) and \cref{fig:Sup:MajoDep}(d.3) for $d_S\lesssim 2 l_B$, i.e., for a separation of the order of the localization of the \gls{ll}, the localized Majorana state leads to the expected perfect \gls{car} conversion~\cite{Clarke2014, Finocchiaro2018}. This way, the hybridized state obtained for $d_S \to 0$ can be interpreted as a delocalized Majorana state with the same quantized \gls{car} signal.

\section{Phase diagrams for other sets of parameters}

\par
\begin{figure}[htb]
    \centering
    \includegraphics[width=0.78\linewidth]{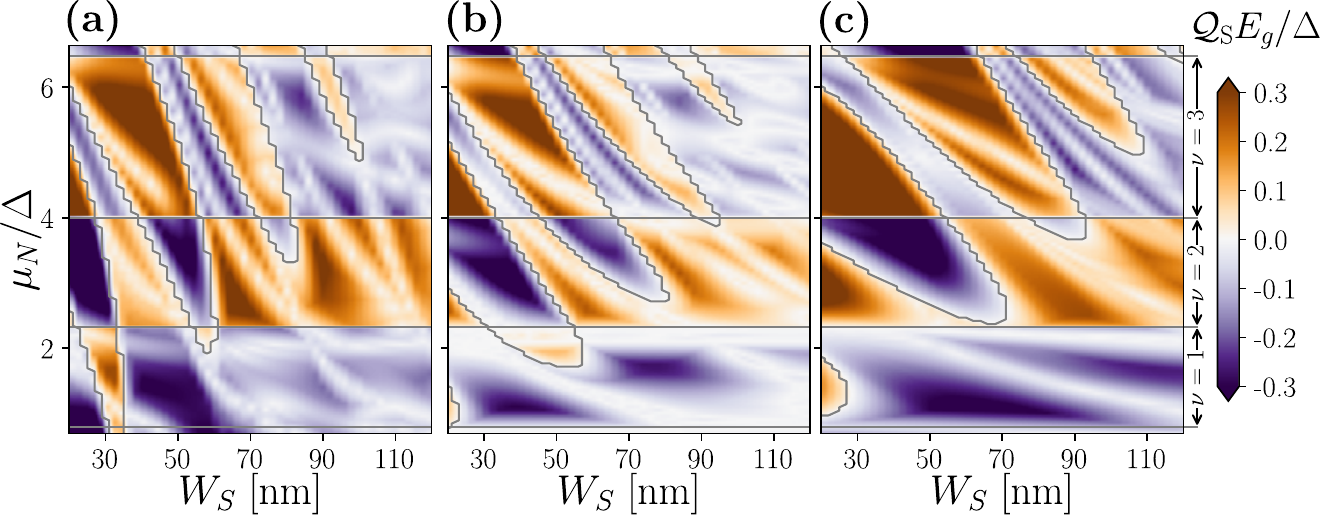}
    \caption{
    Phase diagrams for the same stripe configuration as in \cref{fig:1Sketch}(a) with: (a) Fermi level mismatch in the \gls{sc} proximized region given by $\mu_S/\Delta = 6$; (b) 
    Fermi level mismatch in one of the \gls{qh} regions given by $\mu_N'= \mu_N + \delta\mu$ with $\delta\mu/\Delta = 0.2$; and (c)
    wider penetration depth of the magnetic field and Zeeman splitting ($\lambda_B = 24 ~\si{nm}$).}
    \phantomsection
    \label{fig:Sup:OtherPars}
\end{figure}

 The results presented in the main text and in the End Matter correspond the case of an heterostructure without Fermi level mismatch and with almost perfect magnetic field screening inside the \gls{sc} region ($\lambda_B = 6 \si{nm}$). The phase diagrams are qualitatively similar independently of these conditions, as shown in \cref{fig:Sup:OtherPars}. 

The Fermi level mismatch modifies the evolution of the minigap as a function of $W_S$ and shrinks the topological regions if $\mu_S > \mu_N$, see \cref{fig:Sup:OtherPars}(a) for a fixed value of $\mu_S/\Delta = 6$. 
In \cref{fig:Sup:OtherPars}(b), the case of a broken mirror-symmetric stripe is considered by modifying the chemical potential in only one of the two \gls{qh} regions setting $\delta \mu = 0.2 /\Delta$. Even if the mirror symmetry is protecting the $\mathcal{Q}_{\rm S}=-1$ state, if $\delta \mu < E_g$ the topological phase is still observed. 
Finally, \cref{fig:Sup:OtherPars}(c) considers the case of a wider penetration depth of the magnetic field and Zeeman splitting, parametrized by $\lambda_B = 24 \si{nm}$. In this case the lowest tongue-like region does not reach $\nu=1$ for $W_S > \xi_S$. However, a wide range of stripe widths still shows a topological \gls{sc} gap for higher filling factors. 

\end{document}